\documentclass[fleqn,usenatbib]{mnras}

\usepackage[T1]{fontenc}

\DeclareRobustCommand{\VAN}[3]{#2}
\let\VANthebibliography\thebibliography
\def\thebibliography{\DeclareRobustCommand{\VAN}[3]{##3}\VANthebibliography}

\newcommand{\reffig}[1]{Figure~\ref{#1}}
\newcommand{\refsec}[1]{Section~\ref{#1}}

\usepackage[dvipdfmx]{graphicx}	
\usepackage{amsmath}	
\usepackage{amssymb}	

\usepackage{color}
\usepackage[normalem]{ulem}
\usepackage{xcolor}

\newcommand{\Add}[1]{\textcolor{black}{#1}}	
\newcommand{\Delete}[1]{\if0{#1}\fi}	
\newcommand{\Adds}[1]{\textcolor{black}{#1}}	
\usepackage[version=4]{mhchem}
\usepackage{newtxtext,newtxmath}

\title[Stellar wind effect on the atmospheric escape, max. 45 characters]{Stellar Wind Effect on the Atmospheric Escape of Hot Jupiters \Adds{and their Ly-$\alpha$ and H$\alpha$ transits}}

\author[H. Mitani et al.]{
Hiroto Mitani,$^{1}$\thanks{E-mail: hiroto.mitani@phys.s.u-tokyo.ac.jp}
Riouhei Nakatani,$^{2}$
Naoki Yoshida$^{1,3,4}$
\\
$^{1}$Department of Physics, School of Science, The University of Tokyo, 7-3-1 Hongo, Bunkyo, Tokyo 113-0033\\
$^{2}$RIKEN Cluster for Pioneering Research, 2-1 Hirosawa, Wako, Saitama 351-0198, Japan\\
$^{3}$Kavli Institute for the Physics and Mathematics of the Universe (WPI), UT Institutes for Advanced Study, The University of Tokyo, Kashiwa, Chiba 277-8583, Japan\\
$^{4}$Research Center for the Early Universe, School of Science, The University of Tokyo, 7-3-1 Hongo, Bunkyo, Tokyo 113-0033
}

\date{Accepted XXX. Received YYY; in original form ZZZ}

\pubyear{2021}

\begin{document}
\label{firstpage}
\pagerange{\pageref{firstpage}--\pageref{lastpage}}
\maketitle

\begin{abstract}
Atmospheric escape of close-in exoplanets can be driven by high energy radiation from the host star. 
The planetary outflows interacting with the stellar wind may generate observable transit signals that depend on the strength of the stellar wind. 
We perform \Add{detailed} radiation-hydrodynamics simulations of the atmospheric escape of hot Jupiters with including the wind from the host star in a self-consistent, dynamically coupled manner. We show that the planetary outflow is shaped by the balance between its thermal pressure and the ram pressure of the stellar wind. We use the simulation outputs to calculate the Lyman-$\alpha$ and H$\alpha$ transit signatures. \Delete{The transit peak depth is blue-shifted by $\sim 10\,{\rm km/s}$, but strong}\Add{Strong} winds can confine the outflow and decrease the Lyman-$\alpha$ transit depth. Contrastingly, the wind effect on H$\alpha$ is weak because of the small contribution from the uppermost atmosphere of the planet.
\Add{Observing both of the lines is important to understand the effect of the UV radiation and wind from the host.}
The atmospheric mass-loss rate is approximately independent of the strength of the wind. \Delete{We also discuss the effect of the flare on the transit signature, to conclude that the probability of a flare that would change the Lyman-$\alpha$ transit depth significantly during a single observation is low for solar-type host stars. }\Add{We also discuss the effect of the coronal mass ejections on the signatures. We argue that around M dwarfs the effect can be significant in every transit.}   

\end{abstract}



\section{Introduction} \label{sec:intro}
Recent transit observations detected signatures of atmospheric escape from close-in exoplanets such as hot Jupiters (e.g. HD209458b \citep{Vidal-Madjar_2003}) and hot Neptunes (e.g. GJ 436b \citep{Ehrenreich_2015}). Ultraviolet and X-ray radiation from the host stars can drive the atmospheric escape \citep{Lammer_2003,Yelle_2004, Tian_2005, Murray-Clay_2009, Owen_2012,Tripathi_2015}, which can affect the long-term planetary evolution especially for relatively low-mass and close-in gas giants under weak gravity and strong irradiation \citep{Johnstone_2015,Allan_2019}. 
The escaping atmosphere interacting with the stellar wind generates characteristic Lyman-$\alpha$ transit signals \citep{Etangs_2012,Ehrenreich_2015,Bourrier_2018}. 

Hydrodynamics calculations show that the wind can confine the planetary outflow and reduce the transit depth \citep[e.g.][]{Vidotto_2020, Carolan_2020,Carolan_2021}. The wind effects can be important for young planetary systems in which the host star is active and drives a strong stellar wind. Stellar wind can also suppress the atmospheric escape in close-in planets \citep{Etangs_2012, Garcia_2020}. It is important to study the interplay between the stellar wind and the escaping planetary atmosphere quantitatively by using self-consistent hydrodynamics simulations. Previous theoretical studies examine the dynamical interaction between the stellar wind and the escaping atmosphere \citep[e.g.][]{Carolan_2020,Carolan_2021}, but \Delete{direct simulations of the confined atmosphere with ultraviolet radiation transfer have not been performed} \Add{many of the previous studies of the wind interaction did not follow the launching of the photoevaporative outflows. There have also been studies that do not calculate the structure of the planetary outflow and the wind simultaneously \citep{Christie_2016,Esquivel_2019,Matsakos_2015,Scheiter_2016,Villarreal_2014,Villarreal_2021}.
Photoionization by extreme ultra-violet (EUV) photons is an important process that affects the H$\alpha$ transit depth, which can be observed by ground-based telescopes} \citep{Khodachenko_2019,McCann_2019}. \Add{To understand the observed spectroscopic signatures accurately, it is important to perform radiative transfer calculations and to follow the dynamical interaction of the photoevaporative atmosphere and the stellar wind.}
Multi-dimensional radiation hydrodynamics simulations of the planetary outflow driven by the irradiation of the host star are needed in order to study the detailed atmospheric structure that is confined by a strong stellar wind.

In this {\it Letter}, we perform radiation hydrodynamics simulations of atmospheric escape from a close-in gas giant. We explicitly include the radiation and stellar wind from the host star\Delete{. We} \Add{and} follow the atmospheric escape driven by the ultraviolet photoionization heating, and also its interaction with the stellar wind. Hereafter, unless otherwise mentioned, we use the term `wind' when referring to the stellar wind from the host star, and `outflow' for the escaping atmospheric gas of the planet. 
Our simulations include the radiative transfer of extreme ultraviolet (EUV) photons with $h \nu > 13.6\, {\rm eV}$ and self-consistent non-equilibrium thermochemistry. We run a set of simulations systematically varying the mass-loss rates and EUV flux of the host star.
Finally, we study the observability of the transit signatures caused by time-dependent stellar activities like a flare. 

The rest of the paper is organized as follows. In \refsec{sec:Models}, we present the models of our simulations. In \refsec{sec:Results}, we show the results of our fiducial simulations and the stellar wind strength dependence of the Lyman-$\alpha$ and H$\alpha$ transit signals of our output. In \refsec{sec:Discussion}, we discuss the effects of the time-dependent stellar flare activities \Delete{and the magnetic field on the signals.} \Add{and of the stellar and planetary magnetic fields which are not implemented in our simulations.} 

\section{Numerical Model}\label{sec:Models}
We run 2D axisymmetric simulations by placing a planet at the center of the computational domain. The stellar wind and radiation from the host star are injected from one side of the simulation box.
We use the hydrodynamics simulation code PLUTO \citep{Mignone_2007} suitably modified for our study \citep{Nakatani_2018, Nakatani_2018b}.
Assuming symmetry around the axis parallel to the UV radiation, we adopt 2D cylindrical polar coordinates and solve the following hydrodynamic equations coupled with radiative transfer:
\begin{eqnarray}
\frac{\partial\rho}{\partial t}+\nabla\cdot\rho\vec{v} &=& 0\\
\frac{\partial\rho v_R}{\partial t} +\nabla\cdot(\rho v_R\vec{v}) &=& -\frac{\partial P}{\partial R}-\rho\frac{\partial\Psi}{\partial R}\\
\frac{\partial\rho v_z}{\partial t} +\nabla\cdot(\rho v_z\vec{v}) &=& -\frac{\partial P}{\partial z}-\rho\frac{\partial\Psi}{\partial z}\\
\frac{\partial\rho E}{\partial t}+\nabla\cdot(\rho H\vec{v})&=&-\rho\vec{v}\cdot\nabla\Psi+\rho(\Gamma-\Lambda)
\end{eqnarray}
where $\rho, \vec{v}, P$ are gas density, velocity, pressure. The effective gravitational potential $\Psi$ includes contributions of the star and the planet and also incorporates the centrifugal force due to the orbital motion:
\begin{equation}
    \Psi = -\frac{GM_{\rm p}}{r}-\frac{GM_{*}}{r_{*}}-\frac{1}{2}\frac{GM_{*}r_{*}^{2}}{a^{3}}
\end{equation}
where $a, r_{*}$ represent the semi-major axis and local distance to the host star. \Add{We follow the dynamics of the outflow driven by photoinonization heating. 
The detailed atmospheric thermochemical structure is important in order to study H$\alpha$ transit signatures to which the lower atmospheric layers can contribute.}
We configure plane-parallel stellar radiation originating from negative $z$ direction. We neglect the self-gravity of the atmosphere because the mass of the upper atmosphere is much smaller than the planetary mass. The relevant heating and cooling processes (denoted by $\Gamma$ and $\Lambda$) are described in \refsec{sec:heating_cooling}. We also follow non-equilibrium chemistry
\begin{eqnarray}
\frac{\partial n_{\rm H} y_i}{\partial t}+\nabla\cdot(n_{\rm H} y_i \vec{v})=n_{\rm H} R_i
\end{eqnarray}
where $y_i = n_i/n_{\rm H}$ and $R_i$ represent the chemical abundance and the reaction rate, respectively, with $n_{\rm H}$ being the hydrogen nucleus number density. Our reaction network includes the four chemical species \ce{H, H+, H_2,  e-}.
The ray-tracing radiative transfer of EUV photons is performed in the same manner as in \citet{Nakatani_2018} and \citet{Nakatani_2019}.  

The model parameters are listed in Table~\ref{tab:fid_params}. 
Our fiducial model sets the stellar wind strength in terms of the mass-loss rate that is equal to the solar value $\dot{M}_{\odot}
= 2 \times 10^{-14}\, M_{\odot} {\rm yr}^{-1}$.

\begin{table}
    \centering
        \caption{Model parameters in the fiducial run}
    \begin{tabular}{ll}\hline\hline
        Stellar parameters & \\
        Stellar Mass $M_*$ & $1\, M_{\odot}$ \\  
        Stellar Radius $R_*$ & $1\, R_{\odot}$ \\
        Stellar EUV photon emission rate $\Phi_{\nu}$ &  $1.4\times10^{38}\, {\rm s}^{-1}$ \\
        Stellar wind velocity & $540\, {\rm km/s}$ \\
        Stellar wind temperature & $2\times 10^6\, {\rm K}$ \\
        Stellar wind density & $2.5\times10^{3}\, {\rm g/cm}^3$ \\
        Planetary parameters & \\
        Planet Mass $M_{\rm p}$& $0.3 \, M_\mathrm{J}$\\ 
        Planet Radius $R_{\rm p}$& $1\, R_{\mathrm{J}}$\\
        Semi-major axis $a$ & $0.045\, {\rm AU}$ \\ 
     \hline
    \end{tabular}
    \label{tab:fid_params}
\end{table}

\subsection{Heating and cooling processes}\label{sec:heating_cooling}
Photoionization heating is caused by EUV photons. 
The EUV flux is given by
\begin{equation}
    F_{\nu} = \frac{\Phi_{\nu}}{4\pi a^2}\exp[-\sigma_{\nu}N_{\mathrm{HI}}]
\end{equation}
where $\sigma_{\nu}$ is the absorption cross section as a function of $\nu$ \citep{Osterbrock_2006} and $N_{\mathrm{HI}}$ is the column density of hydrogen atoms.
The photoionization heating rate is given by 
\begin{equation}
    \Gamma_{\rm ph} = \frac{1}{\rho}n_{\mathrm{HI}}\int_{\nu_0}^{\infty} {\rm d}\nu \, \sigma_{\nu} h(\nu-\nu_{0})F_{\nu}
\end{equation}
where $h\nu_0 =  13.6\, {\rm eV}$.  
We adopt the model spectrum of \citet{Fossati_2018} for a $6000\, {\rm K}$ star. The EUV flux weakly depends on the temperature of the star and the difference between our fiducial EUV flux and that of many exoplanet hosting stars is typically within a factor of $\sim3$. Active young stars tend to have large EUV emissivity. We will discuss the effect of increasing/decreasing EUV flux in \refsec{sec:strength}.

Our simulations include hydrogen recombination cooling \citep{Spitzer_1978} and Lyman-$\alpha$ cooling of \ion{H}{1} \citep{Anninos_1997} as major radiative cooling processes. We also note that adiabatic cooling is an efficient cooling process in the atmospheric gas.

\subsection{Initial and boundary conditions}\label{sec:condition}
The computational domain is defined on a region with $R=[0,6]\times10^{10}\, {\rm cm}$ and $z=[-6,6]\times10^{10}\, {\rm cm}$. 
The domain is configured on axisymmetric cylindrical grids with $(N_R, N_z) = (480,960)$.

The initial density profile of the atmosphere is given by a hydrostatic isothermal model
\begin{equation}
\rho(r) = \rho_{\rm p}\exp\left[\frac{GM_{\rm p}}{c_s^2}\left(\frac{1}{r}-\frac{1}{R_{\rm p}}\right)\right],
\end{equation}
where $r$, $\rho_{\rm p}$ and $c_\mathrm{s}$ are the radius measured from the planet center, the density at the surface of the planet ($r=R_{\rm p}$), and the sound speed at the surface, respectively. 
The atmosphere of a hot Jupiter may have a cool lower layer and a hot upper layer \citep{Murray-Clay_2009}. We set the initial temperature of the lower atmosphere ($r<1.1R_{\rm p}$) to $4000\, {\rm K}$ and that of the upper atmosphere ($r>1.1R_\mathrm{p}$) to $10000\, {\rm K}$, with setting the pressure gradient continuous across the boundary between the two layers.  

The \Delete{inner core}\Add{innermost} region of the planet does not dynamically affect the structure of the upper atmosphere. We thus do not follow the hydrodynamics of the inner region ($r<0.85R_\mathrm{p}$) to save the computational cost. 
The temperature and the density there are fixed throughout the simulation, and the outward velocity is set to $v_{\rm out} = 0$, so that there should be no inflow/outflow from the \Delete{core} \Add{innermost} region.

The outer boundary conditions are given by setting the density contrast to be unity at the boundary, to avoid spurious reflections and unphysical accumulation of the escaping gas.  
At the symmetry axis, we adopt conventional reflective axisymmetric boundary conditions.
We note that the planet's atmospheric structure can be, in principle, affected by the physical conditions at the outer boundary. Typically, the upper atmosphere contracts and the temperature decreases when the outflow velocity increases \citep{Tian_2013}. However, in our simulations, the physical properties of the stellar wind, namely its density, temperature and velocity, affect the atmospheric structure more strongly.
We have run some tests and have confirmed that the above somewhat artificial boundary conditions do not significantly change the main results.

\begin{figure*}
    \includegraphics[width=18cm]{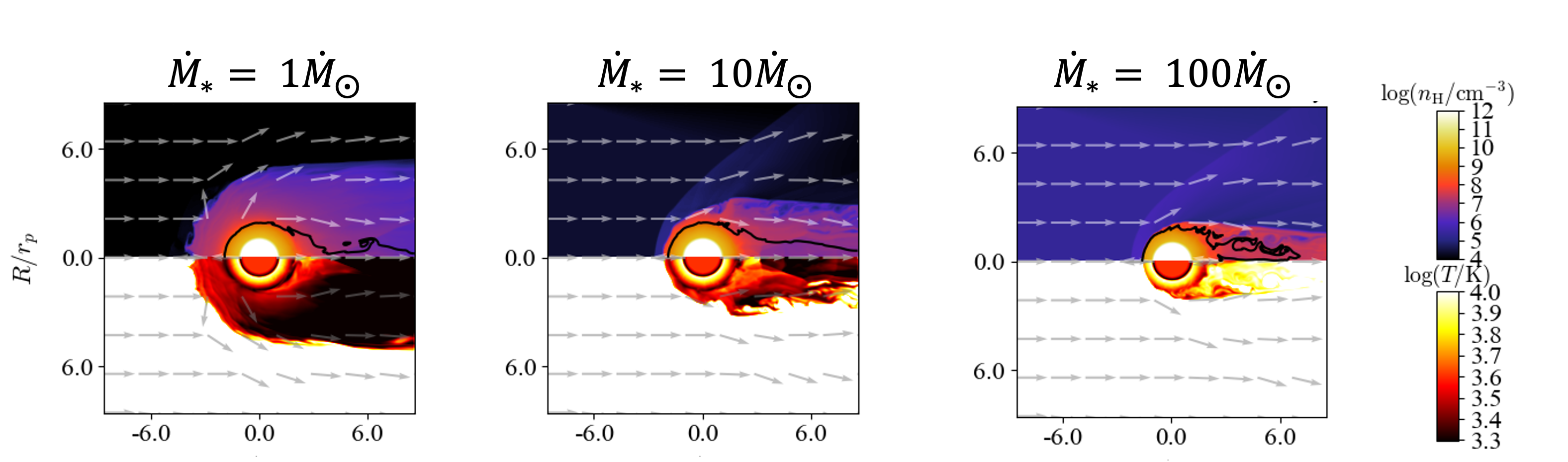}
    \caption{The atmospheric structures in our simulations with $\dot{M_*}=1\,M_{\odot},10\,M_{\odot},100\,M_{\odot}$ at $t=5.8\, {\rm day}$ after the initial conditions. EUV radiation and the wind from the host star are injected from the left side. In each panel, the upper half shows the density distribution and the lower half shows the temperature and the arrows indicate the gas velocity.\Add{The solid lines in upper panels show neutral hydrogen abundance becomes $0.9$. }}
    \label{fig:fid}
\end{figure*}

\section{Results}\label{sec:Results}

\subsection{Fiducial Case}

\reffig{fig:fid} shows the atmospheric structure under the fiducial stellar wind conditions. The temperature of the planetary outflow reaches $\sim 5000\, {\rm K}$ owing to the EUV photoionization, but the outflow is confined by the stellar wind especially in the direction to the host star. \Add{The overall structure appears similar to the previous study which assumed axisymmetry \citep{Christie_2016},
but our radiation transfer calculations reproduce detailed temperature structure
as can be seen in the lower panels.}
The planetary mass-loss rate is as large as $\sim10^{10}\, {\rm g/s}$, which is close to the value without the wind and is similar to the typical mass-loss rates of hot Jupiters. 
Overall, the planetary {\it mass evolution} is not significantly affected by the stellar wind in the typical (our fiducial) cases.

\begin{figure}
    \centering
    \includegraphics[width=9cm]{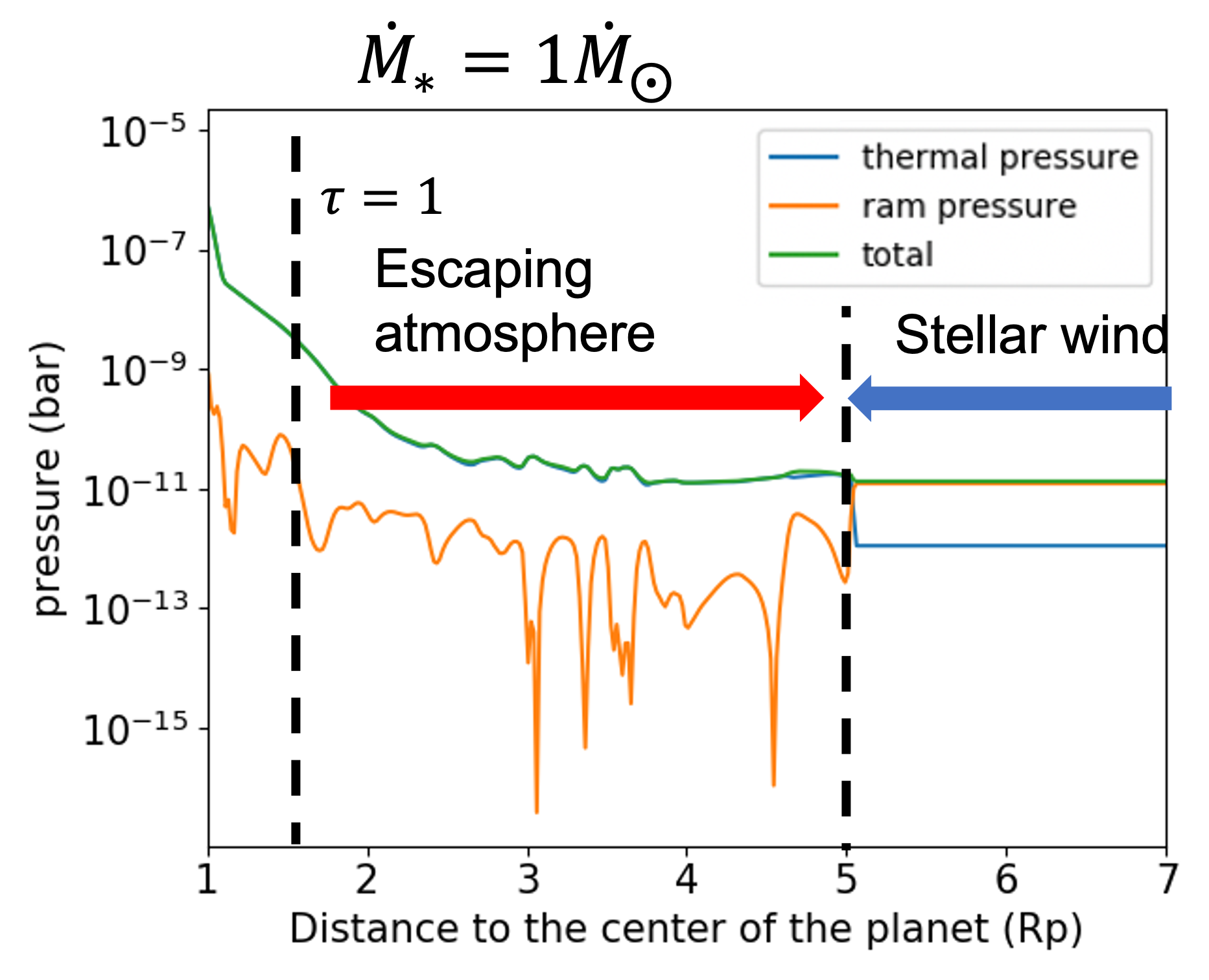}
    \caption{The pressure profiles measured at the symmetry axis in the solar stellar wind case. The thermal pressure of the outflow matches roughly the stellar wind ram pressure at $\sim5 R_{\rm p}$ in solar like stellar wind case. The point where the EUV reaches $\tau=1$ is inner than where the wind confines the outflow.}
    \label{fig:1D}
\end{figure}

The pressure profile is shown in \reffig{fig:1D}. The ram pressure of the wind is matched by the thermal pressure of the outflow at the point $r \sim 5R_{\rm p}$. 
This pressure-balance point can vary depending on the wind property. 
To examine this, we run additional simulations with different wind velocities fixing $\dot{M}_*$. We have found that a higher wind velocity results in a stronger confinement of the atmosphere.
The pressure balance can be expressed approximately as
\begin{equation}
    k_{\rm B}\rho_{\rm p}(r) \, T_{\rm p}(r)/\mu m_{H} = \rho_*(r) \, v_*^2(r)
\end{equation}
where $\rho_{\rm p}(r), T_{\rm p}(r)$ represent the density and temperature of the planetary atmosphere at the contact point, $\mu$ is the mean molecular weight, $m_{H}$ is the hydrogen atomic mass, and $\rho_*(r), v_*(r)$ are the density and velocity of the wind. If we assume spherical symmetry, the mass-loss rate of the planet is
\begin{equation}
\dot{M}_{\rm p}=4\pi r^2 \rho_{\rm p} v_{\rm p},
\end{equation}
and that of the host star is
\begin{equation}
\dot{M}_*=4\pi a^2 \rho_* v_*.
\end{equation} 
Then the effective confinement radius is estimated to be

\begin{equation}
\begin{split}
    r &= \sqrt{\frac{\dot{M}_{\rm p}}{\dot{M}_*}\frac{k_{\rm B} T_{\rm p}}{\mu m_{H} v_{\rm p} v_*}}a\\
    & \approx 4\times10^{10} \, {\rm cm} \left(\frac{\dot{M}_{\rm p}}{3\times10^{10}\, {\rm g/s}}\right)^{1/2}
    \left(\frac{\dot{M}_*}{\dot{M}_{\odot}}\right)^{-1/2} \left(\frac{T_p}{5000\,{\rm K}}\right)^{-1/2}\\
    &\times\left(\frac{v_p}{1\times10^{5}\,{\rm cm/s}}\right)^{-1/2}\left(\frac{v_*}{540\,{\rm km/s}}\right)^{-1/2}\left(\frac{a}{0.045\,{\rm au}}\right) 
\end{split}
\end{equation}

According to this relation, the confinement radius in our fiducial case is a few planetary radii, which is consistent with the pressure-balance point in \reffig{fig:1D}. 

\subsection{Transit signature}
We calculate the Lyman-$\alpha$ and H-$\alpha$ transit depths following the procedure of \citet{Allan_2019}. We use the simulated atmospheric structure and employ a ray-tracing model assuming the Voigt line profile. 
We calculate the transit depth $\delta_{\nu}$ using the optical depth:
\begin{equation}
    \delta_{\nu} = \frac{\int 2 \pi R (1-e^{-\tau_{\nu}}) {\rm d}R}{\pi R_*^2}
\end{equation}
where $\tau_{\nu}$ represents the optical depth of Lyman-$\alpha$/H$\alpha$ at frequency $\nu$. To calculate the H$\alpha$ transit depth, the $n=2$ level population needs to be evaluated because our hydrodynamics simulations do not explicitly follow the level population of hydrogen. 
We calculate the $n=2$ level population using the $2p, 2s$ population of \cite{Christie_2013}:
\begin{equation}
    \begin{split}
    \frac{n_2}{n_1} &= \frac{n_{2p}+n_{2s}}{n_1s} \simeq 10^{-9}\left(\frac{5 R_{*}}{a}\right)^2e^{16.9-(10.2\,{\rm eV}/k_B T_{\mathrm{Ly\alpha}, *})} \\
    &+ 1.627\times10^{-8}\left(\frac{T}{10^4\,{\rm K}}\right)^{0.045}e^{11.84-118400\,{\rm K}/T}\\
    &\times\frac{8.633}{\log(T/T_0)-\gamma}
    \end{split}
\end{equation}
where $T_{Ly\alpha, *} \sim 7000\,{\rm K}$ is the excitation temperature for the solar Lyman-$\alpha$, $T_{0} = 1.02\,{\rm K}$ and $\gamma=0.57721\ldots$ is the Euler-Mascheroni constant. The $2\mathrm{s}$ level population is larger than $n=2\mathrm{p}$ level in the high temperature ($>5000\, {\rm K}$ ) region and $n=2$ abundance is about $10^{-9}$ in the planetary atmosphere of our simulations. In the Lyman-$\alpha$ transit depth, we neglect the $n=2$ level population considering the much greater population of the $n=1$ ground state.

\begin{figure}
    \centering
    \includegraphics[width=8.5cm]{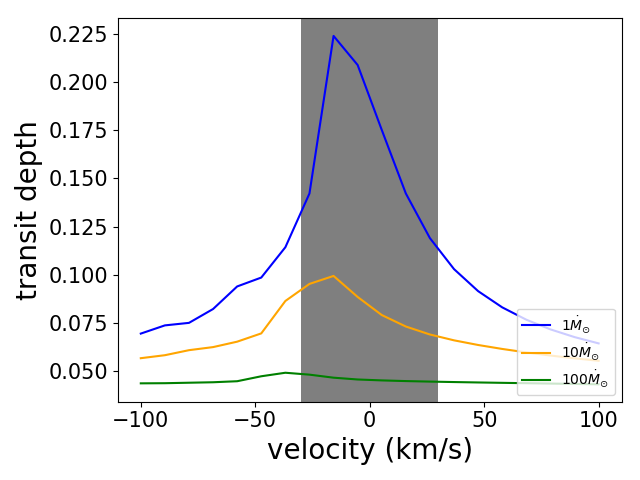}
    \caption{The Lyman-$\alpha$ transit depths for different stellar wind strengths. The shaded region indicates the line center ($-30\,{\rm km/s}<v<30 {\rm km/s}$). The peaks are blue-shifted due to the stellar wind.}
    \label{fig:transit}
\end{figure}

\reffig{fig:transit} shows the resulting Lyman-$\alpha$ transit depth at mid-transit.
The transit depths of Lyman-$\alpha$ are blue-shifted by the stellar wind effect
\Add{, consistent with the findings of, e.g., \citet{Carolan_2020,Carolan_2021}}.  
Note that we have neglected the interstellar absorption, which can be significant around the line center (the shaded region in \reffig{fig:transit}). In practice, one needs to know the relative velocity between the system and the local interstellar medium to predict the observed signal \citep[e.g.,][]{Dring_1997}. 

\begin{figure}
    \centering
    \includegraphics[width=8.9cm]{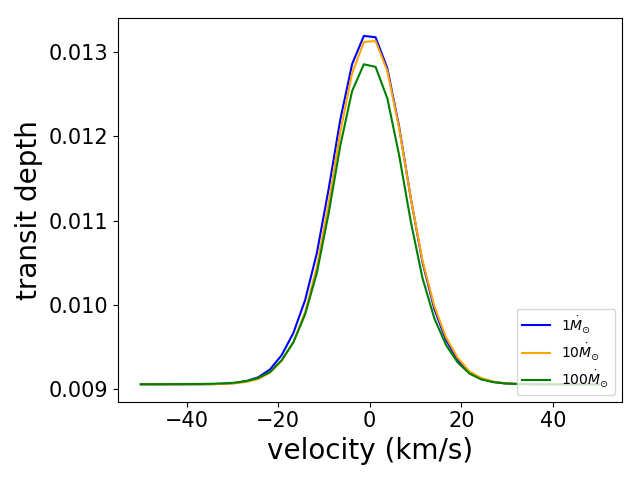}
    \caption{The H$\alpha$ transit depths calculated using our simulation outputs. }
    \label{fig:Halpha}
\end{figure}

\reffig{fig:Halpha} shows that the H$\alpha$ transit depth is smaller because of the small fraction of hydrogen at the $n=2$ level and absorption in the lower region is more important in the H$\alpha$ case than Lyman-$\alpha$ transit depth due to the high temperature in the lower region. 
In general, H$\alpha$ transit observations have advantages to be performed by ground-based telescopes, even though the transit depth is smaller; Lyman-$\alpha$ can be observed only by space telescopes such as Hubble Space Telescope and its line center is affected by interstellar medium absorption \citep{Allan_2019}. 

\subsection{Stellar wind strength and the implication for observations}\label{sec:strength}
We run simulations with different stellar wind strengths (\reffig{fig:fid}). We vary the density of the stellar wind while fixing the velocity and the wind temperature; the velocity structure depends mainly on temperature. The wind ram pressure is proportional to the density of the wind, but the planetary mass-loss rate is insensitive to the wind strength 
(Table~\ref{tab:mass-loss}). The stellar wind largely determines the structure of the escaping atmosphere\Add{; the confinement radius can be estimated as explained in Section 3.1.
Strong stellar winds can reduce the Lyman-$\alpha$ transit depth due to the confinement by the strong ram pressure of the wind \citep[see also][]{Christie_2016,Carolan_2021}}. 

We also run a simulation of an extremely strong stellar wind case of $\dot{M}=1000\,\dot{M}_{\odot}$. In this run, the strong wind can confine the planetary atmosphere within the launching point, and the atmospheric escape is effectively prevented. In such an extreme case, the planetary mass evolution can be largely determined by the stellar wind rather than the radiation-driven atmospheric escape.

Observationally, the absence of a clear Lyman-$\alpha$ signal may indicate a very strong stellar activity in progress. A strong stellar wind ($> 100\,\dot{M}_{\odot}$) may be expected for young stars ($\sim100 {\rm \, Myr}$), but such stars would also have a high EUV emission rate, and thus tend to enhance the transit depth. Young stars may thus have a clearer signal than predicted by the $100\,\dot{M}_{\odot}$ case in \reffig{fig:transit}.

Recent observations suggest stellar winds with $\dot{M}\sim30\,M_{\odot}$ for active M~dwarfs \citep{Wood_2021}. It implies that even for young stars ($\sim100 {\rm \, Myr}$), there may be many close-in exoplanets showing noticeable Lyman-$\alpha$ transit signals.

\begin{table}
    \centering
        \caption{Planetary Mass-loss rates with different stellar winds}
    \begin{tabular}{ll}\hline\hline
        Stellar Wind strength & Mass-loss rate (g/s)\\
        $1\,\dot{M}_{\odot}$ & $2.9\times10^{10}$ \\
        $10\,\dot{M}_{\odot}$ & $2.2\times10^{10}$ \\
        $100\,\dot{M}_{\odot}$ & $2.3\times10^{10}$ \\
     \hline
    \end{tabular}
    \label{tab:mass-loss}
\end{table}

As discussed in the above, a very strong stellar wind can confine the upper atmosphere and reduce the Lyman-$\alpha$ transit depth. 
We also run simulations with different EUV flux and find that intense EUV radiation drives strong outflows and increases the transit depth. Thus the effects of the stellar wind and EUV radiation degenerate in terms of the Lyman-$\alpha$ transit depth.

In \reffig{fig:Halpha}, we show the transit depth of H$\alpha$ for various stellar wind strengths. We find that the effect of the wind confinement is not significant in H$\alpha$ transit because the contribution from the low temperature and low density region is small (\reffig{fig:Halpha}). To reduce the H$\alpha$ transit depth by the wind confinement, a very strong wind with $\dot{M}_*>1000\,\dot{M}_{\odot}$ is needed. Strong EUV heats the gas and increases the H $\alpha$ transit depth due to the large population of $n=2$ level.  
The qualitative difference of the wind effect between Lyman-$\alpha$ and H$\alpha$ suggests that we can observe the H$\alpha$ transit around active stars. We can also estimate the stellar wind strength using Lyman-$\alpha$ transit depth. Such a diagnostics can be used to break the degeneracy of the stellar wind strength and EUV flux in transit. 

\Add{Previous 3D simulations argue that the tail of the outflow contributes to the Lyman-$\alpha$ blue-wing. We estimate the contribution in the following manner.
The tail scale length would be 
\begin{equation}
    R_t = \frac{u}{\Omega}
\end{equation}
where $u$ is the velocity of the outflow and $\Omega$ is the angular velocity of the planet. The ratio $R_t$ is typically a few planetary radii, which can be large enough to absorb the Lyman-$\alpha$ from the star. We assume that the tail size is $R_t$ and the impact parameter $b=0$. We estimate the density of the tail to be 
similar to the density behind the planet in our simulations, and the ionization fraction can be estimated assuming the ionization equilibrium. The Lyman-$\alpha$ cross section is $\sim8\times10^{-18}\,{\rm cm^{2}}$ and the density of the tail is $\sim10^{6}\,{\rm cm^{-3}}$. The tail is nearly fully ionized with $X_{\rm HI} \sim 10^{-2}$ because of the intense EUV radiation. 
\citet{Villarreal_2021} concluded that the amount of neutral material in the tail is negligibly small in the case of warm Neptunes with strong UV irradiation, whereas \citet{Carolan_2021} use a 1D model to estimate the neutral density and argue that the comet-like tail causes line profile asymmetry. This clearly warrants further study using 3D simulations with radiative transfer.
We note that, in the case of H$\alpha$, 
the $n=2$ level population is extremely small in the thin tail (Equation 15) and thus the tail does not significantly contribute to the transit depth.
}

\section{Discussion}\label{sec:Discussion}
\subsection{The effect of time-dependent stellar activities on the outflow}
Stellar activities drive winds and the strength can be time-dependent. In the case of the sun, coronal mass ejections
\Add{(CMEs)} and associated flare activities are well-known. \Add{Strong CMEs can confine the outflow and change the absorption signals.}
\Delete{The occurrence rate depends on the strength of the activity.}
\Add{The strong flares are accompanied by CMEs in many cases for the sun. The frequency of the CMEs can be estimated by the solar flare frequency.}
The frequency of strong flares ($>10^{34}\,{\rm erg}$) is very low with an approximate rate of $<10^{-34}\, {\rm year}^{-1} {\rm erg}^{-1}$ \citep{Maehara_2017}, and thus the possibility to observe transit during a strong flare is likely small. For giant flares with the frequency of 1 per year, the possibility to observe a transit with duration of a few hours is less than $0.1 \%$. According to the observed relation between flare energy and coronal mass ejection rate \citep{Drake_2013}, the mass-loss rate is about 10 times higher when the flare energy is $\sim10^{32}\, {\rm erg}$ . 
If we focus on the flare frequency $f(E_{\rm flare})$ corresponding to the flares which can confine the upper atmosphere and can reduce the transit depth ($E_{\rm flare}>10^{32}\,{\rm erg}$), the occurrence rate is estimated by using the power-law relation $f(E_{\rm flare})\propto E_{\rm flare}^{-2}$\citep{Maehara_2017},
\begin{equation}
    \int_{10^{32}\,{\rm erg}}^{\infty} f(E_{\rm flare}) \,{\rm d}E_{\rm flare}\sim 1-100\, {\rm year}^{-1}
\end{equation}
and the possibility to observe transit during a flare is less than $1\%$ even if we consider all the flare activities which can cause an order of magnitude larger mass-loss than the sun. \Add{This probability is for the solar type stars. The CME activity may be smaller in the other stars\citep{Leitzinger_2014} but the activity can be significant in very young active systems. Young stars are active and their strong flare frequency is larger than the old stars  \citep{Feinstein_2020}. This may be important only if the age is below $50\,{\rm Myr}$ when the difference in frequency is large.
The frequency of the strong flares also depends on the spectral type of the stars. The rate in M dwarfs is a few orders magnitude larger than in G-type stars. This suggests that there are frequent strong flares around M dwarfs and that the Lyman-$\alpha$ transit is affected by CMEs almost every transit. Around K and M-type stars, the Lyman-$\alpha$ transit can be reduced by the strong stellar activities, but the H$\alpha$ transit may be observed (\reffig{fig:Halpha}).
}
\Adds{We note that, for active host stars, it would be difficult to disentangle the transit signature from that of the background stellar activity. This can be a potential problem in observations of H$\alpha$ signals.}
\subsection{Magnetic fields effect on the outflow}
The stellar wind is highly ionized and can interact with the planetary magnetic field. The strength and structure of the magnetic field of an exoplanet are not well understood, and it may likely depend on the properties of the planet and the host system. In the case of rocky planets, the pressure of the planetary magnetic field is smaller than the thermal pressure of the outflow \citep{Zhang_2021}. \Add{Previous studies use magneto-hydrodynamics simulations of hot Jupiters 
to examine the effect of the planetary magnetic field \citep{Matsakos_2015,Carolan_2021b,Odert_2020,Villarreal_2018}.} We can estimate the effects of planetary magnetic fields using the ratio of the pressure of outflow to the magnetic pressure. For a magnetic field with $\sim10\, {\rm G}$, the magnetic pressure is $P_{B} \sim 1\, \mu{\rm bar}$ . If the hot Jupiters have a strong magnetic field, it can shield the entire atmospheric outflow against the wind. Detailed simulations with radiative transfer are needed to evaluate the wind effect with the strong planetary magnetic field, as suggested by \citet{Cauley_p_2019}.

\Add{More recently, \citet{Harbach_2021} study the effect of the magnetized wind on the absorption signatures. 
In \refsec{sec:Results}, we consider only the hydrodynamic ram pressure of the wind to derive the confinement condition. 
The interaction of the magnetized wind and the planetary outflow is highly complex and variable \citep{Harbach_2021}, and thus further systematic study would be needed.}

\section*{Acknowledgements}
HM has been supported by International Graduate Program for Excellence in Earth-Space Science (IGPEES) of the University of Tokyo and JSPS KAKENHI Grant number 21J11207.
RN is supported by Grant-in-Aid for Research Activity Start-up (19K23469) and the Special Postdoctoral Researcher program at RIKEN.
The numerical computations were carried out on Cray XC50 at Center for Computational Astrophysics, National Astronomical Observatory of Japan.

\section*{Data Availability}
The data used in this paper will be shared upon request to an author of this paper.

\bibliographystyle{mnras}
\bibliography{atmospheric_escape}


\bsp	
\label{lastpage}
\end{document}